\documentclass[letterpaper]{article}

\usepackage[preprint]{preprint}
\usepackage[hyphens]{url}
\usepackage{graphicx}
\usepackage{natbib}
\usepackage{caption}
\usepackage{algorithm}
\usepackage{algorithmic}
\usepackage{newfloat}
\usepackage{listings}
\DeclareCaptionStyle{ruled}{labelfont=normalfont,labelsep=colon,strut=off}
\floatstyle{ruled}
\newfloat{listing}{tb}{lst}{}
\floatname{listing}{Listing}

\usepackage{booktabs}
\usepackage{multirow}
\usepackage{amsmath}

\title{When Memory Takes Gradients: Collaborative Vector Memory for Agentic Recommender Systems}
\author{
  Hanchong Chen,
  Xing Tang,
  Lingjie Li,
  Xiongfeng Shan,
  Xiuqiang He
}
\affiliations{
  Shenzhen Technology University\\
  hanchongchen.cs@gmail.com, xing.tang@hotmail.com, lilingjie@sztu.edu.cn,\\
  xiongfeng.shan@hotmail.com, he.xiuqiang@gmail.com
}

\begin{document}

\maketitle

\begin{abstract}
Agentic recommender systems ground each decision of a large language model (LLM) in a persistent memory of the user, and in existing agents that memory is text: a narrative written and maintained by further LLM calls. Text limits this memory in two ways. It is updated one rewrite at a time, so exploiting the full interaction history is prohibitively expensive; and collaborative evidence, graded similarity over an entire catalog, does not survive translation into sentences. We propose CoVeMem (Collaborative Vector Memory), which vectorizes the collaborative core of the agent's memory. Frozen LightGCN user and item states form the memory bank; at each decision, the candidate set itself retrieves the most relevant historical states, which enter the LLM's context as soft tokens alongside a light textual profile. Contrastive alignment to item-semantic anchors, followed by listwise co-training with masked candidates, teaches the model to read these states and to rank through them; a pointwise yes/no readout scores each candidate. Across four instruction-grounded recommendation benchmarks, CoVeMem matches or exceeds the strongest collaborative text-memory agent on 19 of 20 metric cells while requiring zero additional LLM calls for memory maintenance beyond the shared static profile, against per-interaction calls for text memory. The memory now takes gradients: the full interaction history, out of reach for text, becomes available as training data for what the agent remembers and for how it reads what it remembers.
\end{abstract}

\section{Introduction}
\label{sec:intro}

Recommendation has recently seen a growing line of work that hands the decision to an LLM agent \citep{zhao2024llmrecsurvey,peng2025agentsurvey,zhang2025agentirsurvey}. Agentic recommender systems place such an agent between the user and the recommendation platform, where it maintains a persistent memory of the user and adjudicates the platform's candidates on the user's behalf \citep{xu2025iagent,memrec2026}. Within this paradigm, memory is what lifts the agent above stateless prompting \citep{liu2023chatgpt}. It accumulates what the agent learns about the user across sessions and carries the taste that the current instruction leaves unsaid. How well the agent serves the user therefore depends not only on what its memory retains, but also on how its contents are represented and made available to the LLM.

\begin{figure}[t]
  \centering
  \includegraphics[width=\columnwidth]{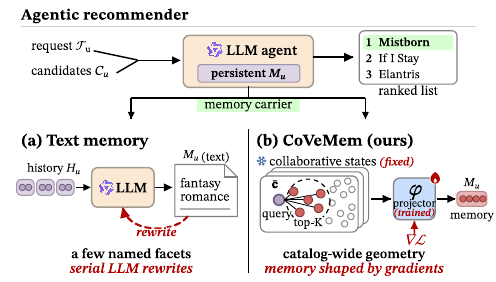}
  \caption{(a) Existing agents maintain a text narrative through serial LLM rewrites, so the memory keeps only what it names. (b) CoVeMem keeps catalog-wide collaborative states and learns to read them from ranking gradients, so the full interaction history becomes training data.}
  \label{fig:teaser}
\end{figure}

In recent work, the agent's persistent memory has almost always been represented as text. As illustrated in Figure~\ref{fig:teaser}(a), the interaction history is distilled into an evolving user narrative through serial LLM rewrites, and this textual memory is subsequently read back into the agent's prompt at decision time. Existing systems have expanded both what this narrative contains and how it evolves, from static profiles \citep{xu2025iagent} and self-reflective updates \citep{zhang2024agentcf} to collaborative facets \citep{memrec2026}, personalized policy skills \citep{sager2026}, and memory coupled with tool use \citep{tang2025recbot}. Yet their common textual form imposes two fundamental limits. First, the memory is maintained serially, one LLM rewrite at a time. Absorbing each new interaction requires another call, making it expensive and slow to exploit the full interaction history. Second, collaborative structure is compressed when translated into sentences. MemRec propagates writes along the interaction graph and distills neighboring-user facets into the narrative \citep{memrec2026}, while AgentCF co-adjusts user and item memories after each interaction to imitate collaborative filtering in text \citep{zhang2024agentcf}. Yet such memories can name only a few neighbors or facets at a time, whereas the underlying collaborative structure consists of graded relations spanning an entire catalog. Much of this geometry is lost during serialization, because text carries only what it names. Existing work therefore leaves open how to build a persistent agent memory that preserves catalog-wide collaborative structure, can be trained from the full interaction history, and remains directly usable by the LLM without recurring textual rewrites.

Existing text memories can be rewritten from feedback, but ranking gradients cannot directly update the stored text. We therefore make the collaborative memory system trainable by recommendation objectives. Interaction data can then shape the stored user and item states, while ranking losses optimize the way the LLM reads them. The central challenge is to make such a memory usable by the LLM. Collaborative states are not LLM tokens, and the full state bank is too large to fit into the prompt. The agent must first retrieve the states most relevant to the current candidate set and then translate them into a form the LLM can interpret. Even after this translation, the model may still rely on candidate titles and ignore the collaborative states.

We address these problems with CoVeMem (Collaborative Vector Memory), illustrated in Figure~\ref{fig:teaser}(b). CoVeMem learns the contents of its collaborative memory from the interaction graph and learns how the LLM reads that memory from the ranking task. LightGCN user and item states form the memory bank \citep{he2020lightgcn}, while a short textual profile retains explicit user information. For each decision, the candidate set retrieves the historical states most relevant to the current choice. A projector maps the retrieved history, user, and candidate states into soft tokens. The projector and a LoRA adapter together form the trainable read interface of the memory, which we call the \emph{parametric memory reader}. We train this reader through semantic alignment followed by masked listwise co-training. Candidate masking creates comparisons that cannot be resolved from titles alone, so the model must learn to rank through the collaborative tokens. At inference, the agent reads both forms of memory and scores each candidate with a pointwise yes/no query. Across four instruction-grounded benchmarks from InstructRec \citep{xu2025iagent}, CoVeMem exceeds the strongest text-memory baseline in nearly all evaluated settings, with especially clear gains in interaction-dense domains. It also avoids recurring LLM calls for memory maintenance beyond the one-time textual profile.

Our contributions are summarized as follows:
\begin{itemize}
    \item We identify fundamental limitations of text-based memory in agentic recommendation. Serial LLM rewrites make memory costly to maintain, and ranking gradients cannot directly update the stored text. Textual serialization also loses catalog-wide collaborative structure.

    \item We propose CoVeMem, a collaborative memory built from graph-trained user and item states. It retrieves history relevant to each candidate set and maps these states into the LLM through its parametric memory reader, formed by the projector and LoRA adapter. Masked listwise co-training creates comparisons that titles alone cannot resolve and trains the model to rank through the memory.

    \item We conduct extensive experiments on four InstructRec benchmarks. CoVeMem exceeds the strongest text-memory baseline in nearly all evaluated settings, with especially clear gains in interaction-dense domains. It also avoids recurring LLM calls for memory maintenance.
\end{itemize}

\section{Related Work}
\label{sec:related}

Two lines of research are most relevant to CoVeMem. The first studies memory in LLM agents for recommendation. The second integrates collaborative representations into LLM recommenders.

\textbf{LLM agents for recommendation.}
A growing body of work applies LLM agents to recommendation \citep{peng2025agentsurvey}. Memory is central to these agents because it carries past preferences and interactions into later decisions. iAgent grounds recommendations in a static textual profile, while i$^2$Agent maintains a dynamic textual memory updated from interaction feedback \citep{xu2025iagent}. AgentCF co-adjusts user and item text memories after each interaction to approximate collaborative filtering \citep{zhang2024agentcf}. RecBot couples memory with tool use \citep{tang2025recbot}. MemRec introduces collaborative signals by retrieving neighboring-user facets and propagating text-memory updates along the interaction graph \citep{memrec2026}. Sager further distills personalized policy skills on top of an evolving textual memory \citep{sager2026}. Across these systems, collaborative signals may affect how memory is constructed, but the persistent memory read by the LLM remains text. MemRec also considers a vector variant, but this variant delegates ranking to vector similarity rather than teaching the LLM to read vector states as memory \citep{memrec2026}. CoVeMem instead represents the collaborative part of persistent agent memory with vector states and optimizes how the LLM reads those states through ranking supervision.

\textbf{Integrating collaborative information into LLMs.}
A parallel line of work injects collaborative representations into LLMs for conventional recommendation. CoLLM projects user and item representations from a collaborative model into the LLM token space with an MLP and adapts the LLM through LoRA \citep{zhang2025collm}. LLaRA pairs behavioral tokens with item text \citep{llara2024}, while E4SRec represents an interaction sequence with one soft token for each item and scores over the full catalog \citep{li2023e4srec}. \citet{hossain2025embeddings} provide the LLM with a user embedding and multiple historical-item embeddings, using projector training followed by joint adaptation with LoRA. SAILRec keeps its collaborative embeddings frozen, aligns them with textual semantics, adapts the LLM with LoRA, and scores candidates through a yes/no readout \citep{sailrec2026}. To the best of our knowledge, CoVeMem is the first to use collaborative representations as the persistent collaborative component of an agentic recommender's complete memory, combining a persistent state bank, candidate-conditioned retrieval, and a ranking-trained parametric memory reader rather than treating the representations only as conditioning inputs to a conventional LLM ranker.

\section{Method}
\label{sec:method}

\textbf{Problem formulation.}
Let $\mathcal{U}$ and $\mathcal{I}$ denote the sets of users and items. Each user $u$ carries a chronologically ordered interaction history $H_u = [i_1, \dots, i_T]$. At decision time the system receives three inputs: the user, a free-text \emph{instruction} $\mathcal{T}_u$ stating what the user currently wants (a constraint, a preference, a goal), and a candidate set $C_u \subset \mathcal{I}$ containing one ground-truth item among sampled negatives. The task is to rank the ground truth first. We work in the instruction-grounded InstructRec setting \citep{xu2025iagent}, whose datasets ship one instruction per test case; splits and candidate construction follow \S\ref{sec:setup}.

\textbf{Overview.}
An agentic recommender answers this task with an LLM $\mathcal{M}$ grounded in a persistent per-user memory $M_u$, rendered as the complete memory context of each decision prompt. Existing agents implement $M_u$ entirely in natural-language text around the LLM, at the structural costs identified in the introduction. We vectorize the collaborative component of the agent's memory while retaining a light textual profile. At each decision, $M_u$ combines this profile with collaborative user, history, and candidate states drawn from a memory bank trained over the interaction graph.

\begin{figure*}[t]
  \centering
  \includegraphics[width=0.92\textwidth]{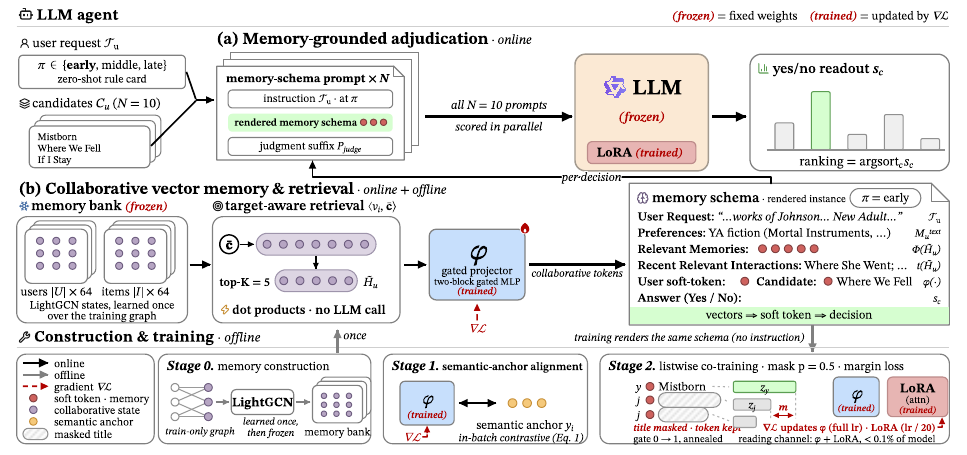}
  \caption{CoVeMem's hybrid memory system: the collaborative vector memory stores graph-trained user and item states, and the parametric memory reader lets the LLM read them to score each candidate.}
  \label{fig:pipeline}
\end{figure*}

Two components make this a hybrid memory system: the collaborative vector memory, which stores the evidence, and a parametric memory reader, which learns to use it. The reader, a gated projector and a LoRA adapter shared across all users, encodes how the injected states are translated into the language space and exploited for ranking. The memory therefore takes gradients not only when its contents are trained over the interaction graph, but also when the reader learns how those contents should be read. The rest of this section follows Figure~\ref{fig:pipeline}: the collaborative vector memory (\S\ref{sec:construct}), learning to read it (\S\ref{sec:train}), and memory-grounded adjudication (\S\ref{sec:adjudicate}).

\subsection{Collaborative Vector Memory}
\label{sec:construct}
\label{sec:read}

\textbf{Memory bank.}
The memory bank stores one $d$-dimensional state per user and per item ($d{=}64$), $\{\mathbf{s}_u\}_{u \in \mathcal{U}} \cup \{\mathbf{v}_i\}_{i \in \mathcal{I}}$, learned once over the bipartite graph of training interactions with standard LightGCN propagation \citep{he2020lightgcn} and then frozen. The graph is built from training interactions only, so no validation or test signal reaches the states. Each user also keeps a short textual profile $M_u^{\mathrm{text}}$, distilled once from training reviews, following common practice in agentic recommenders.

For the LLM to make use of the collaborative information, the relevant states must reach its context at each decision.

\textbf{Target-aware retrieval.}
For a decision over candidates $C_u$, the candidate centroid $\bar{\mathbf{c}} = \tfrac{1}{|C_u|}\textstyle\sum_{c \in C_u} \mathbf{v}_c$ queries the history available at that decision, $\tilde{H}_u = \operatorname*{arg\,top\text{-}K}_{i \in H_u^{\mathrm{avail}}} \langle \mathbf{v}_i, \bar{\mathbf{c}} \rangle$, returning the $K$ historical items ($K{=}5$) whose states are most relevant to the present choice, rather than the most recent ones. The available history $H_u^{\mathrm{avail}}$ is the user's training sequence at inference and the strict causal prefix of the event at training (\S\ref{sec:train}); in both cases the item being scored lies outside it, so retrieval can never surface the target. It is candidate-conditioned retrieval in state space, and it requires only a few dot products and no LLM call.

\textbf{Memory injection.}
A gated projector $\phi$ (a two-block gated MLP; architecture in the technical appendix) maps a state into the input-embedding space of $\mathcal{M}$, where it occupies a single token position as a \emph{soft token}. We name these soft tokens by their roles---the user token, item tokens, and token-history tokens---and refer to them collectively as \emph{collaborative tokens}. The projector is the only path by which the state vectors enter the model; \S\ref{sec:train} trains it. The prompt then carries a fixed \emph{memory schema}: the user's textual profile, a \emph{token-history} line with exactly $K$ slots holding $\phi(\mathbf{v}_i)$ for $i \in \tilde{H}_u$, a titles line naming the same $K$ events, the user token $\phi(\mathbf{s}_u)$, and the candidate lines pairing each title with its item token $\phi(\mathbf{v}_c)$.
The titles line is a light textual complement to the token-history (\S\ref{sec:soft-token-ablation} separates the two).

The bank and profile persist across decisions; the selection is recomputed for each decision. A remaining problem is that $\mathcal{M}$ cannot yet read the collaborative tokens: they are vectors, not language. Section~\ref{sec:train} teaches the model to read and use them.

\subsection{Learning to Read the Memory}
\label{sec:train}

Training bridges the modality gap between the collaborative states and the language space. The stored states stay frozen; what gradients train is the parametric memory reader: the projector $\phi$, which translates the states into the language space, and a rank-$r$ LoRA adapter \citep{hu2022lora} on the attention projections, which adapts how attention reads them, together about 6.8M parameters, under $0.1\%$ of the model. The reader is procedural: it stores how collaborative evidence is read, not what any one user prefers. User-specific content stays in the memory bank and the profile.
Training proceeds in two stages: alignment gives every projected state a meaning, and co-training makes that meaning matter for ranking.

\textbf{Stage 1: semantic-anchor alignment.}
For each item $i$, we construct a semantic anchor $\mathbf{y}_i$ by averaging the LLM input embeddings of its title and available category tokens, using the description when category information is unavailable. This provides free supervision for the modality bridge. We train $\phi$ with a symmetric in-batch contrastive loss \citep[cf.][]{radford2021clip},
\begin{equation}
\mathcal{L}_{\mathrm{align}}
= \tfrac{1}{2}\left[\mathrm{CE}\!\left(\tfrac{\hat\phi(\mathbf{v})\,\hat{\mathbf{y}}^{\!\top}}{\tau}\right)
+ \mathrm{CE}\!\left(\tfrac{\hat{\mathbf{y}}\,\hat\phi(\mathbf{v})^{\!\top}}{\tau}\right)\right],
\label{eq:align}
\end{equation}
where $\hat{\cdot}$ is $\ell_2$ normalization, $\tau$ is a temperature, and targets are the diagonal. The contrastive form forces each projected state to discriminate its own item within the batch. This stage serves as initialization: it places projected states in meaningful regions of the language space, and Stage 2 builds on it.

\textbf{Stage 2: listwise co-training.}
Every training interaction with a sufficient prefix becomes a ranking event: the prefix is the history, the interacted item is the positive, and $N{-}1$ negatives are drawn outside the user's entire sequence. The event's history, its retrieved titles, and its candidates are strictly prefix-causal: the positive and any later interaction are excluded from the history and enter only as a (possibly masked) candidate. Two components are static by design---the memory bank and the textual profile are estimated once over the full training period (\S\ref{sec:construct},~\S\ref{sec:setup}) and held fixed across a user's events, so an early event's states may already reflect later training interactions. The text-memory baselines make the same transductive use of the training period under the same leave-one-out split. The test target is held out from both the graph and the profile, so the reported test metrics remain disjoint from the memory. Because the carrier is trainable, all of these events become supervision. Each event is rendered with the memory schema of Section~\ref{sec:read} and scored listwise through option logits $z_1, \dots, z_N$; Figure~\ref{fig:pipeline} (stage 2) shows the masked rendering. The listwise form is deliberate: it places the candidates in direct competition within one context, so once masking (below) removes the textual route, the ranking loss can only be lowered by comparing the collaborative tokens against one another, which is what forces the model to learn to read them.
One component is deliberately absent from training: the instruction. InstructRec's instructions are derived from each user's held-out target item and therefore constitute target-conditioned information. They are unsuitable for the causal training protocol used here, so instructions enter only at inference. Except for the instruction, training and inference prompts carry the same components.
\textbf{Preventing textual shortcuts.}
Without further constraints, this objective ignores the memory: candidate titles alone carry enough signal to rank by text. Two coupled designs suppress the shortcut. First, candidate masking hides each candidate's title with probability $p{=}0.5$; on a masked candidate, the item token becomes the candidate's identity. Second, the loss is a margin objective over a competition pool $\mathcal{P}$ chosen to keep information symmetric: when the positive is masked, $\mathcal{P}$ is the set of masked candidates, so these rows cannot be resolved from titles and must rely on the collaborative tokens; when the positive is unmasked, $\mathcal{P}$ is the full candidate set:
\begin{equation}
\mathcal{L}_{\mathrm{rank}}
= \frac{1}{|\mathcal{P}|-1} \sum_{j \in \mathcal{P},\, j \neq y} \max\!\big(0,\; m - (z_y - z_j)\big),
\label{eq:rank}
\end{equation}
where $y$ indexes the positive, $m$ is the margin, and the pool must hold a competitor ($|\mathcal{P}| \ge 2$).

Two measures stabilize training. The collaborative tokens are scaled by a gate annealed from $0$ to $1$ over the first $T_g$ steps and fixed thereafter, opening the memory channel only as the projector becomes reliable, and light dropout on the user token and the history slots prevents overfitting to any fixed layout position.

\subsection{Memory-Grounded Adjudication}
\label{sec:adjudicate}

The memory is built and the model reads it; what remains is the decision itself. The agent renders the working memory, inserts the instruction, and scores the candidates.

\textbf{Candidate scoring.}
At inference the collaborative signal is a few soft tokens among hundreds of text tokens. A single listwise prompt amplifies this imbalance: the descriptions of all competing candidates fill the context, and the verdict depends on generating and parsing a ranked list. We therefore score candidates pointwise with a yes/no readout: each candidate gets its own compact prompt, in which its collaborative tokens stay salient; all $N$ prompts are scored in parallel; and no text is generated.
The prompt is the memory schema of Section~\ref{sec:read} carrying that candidate alone, plus the instruction inserted at a dataset-level position $\pi$ (below); the score is the logit of the ``Yes'' token at the answer position,
\begin{multline}
s_c = \mathcal{M}\big(\mathcal{T}_u \,\Vert_{\pi}\, \big[\, M_u^{\mathrm{text}} \,\Vert\, \Phi(\tilde{H}_u) \,\Vert\, t(\tilde{H}_u) \\
\Vert\, \phi(\mathbf{s}_u) \,\Vert\, (c,\, \phi(\mathbf{v}_c)) \,\Vert\, P_{\mathrm{judge}} \,\big]\big)\big[\texttt{Yes}\big],
\label{eq:score}
\end{multline}
where $\Vert$ denotes concatenation ($\Vert_{\pi}$: with the instruction inserted at position $\pi$), $\Phi(\tilde{H}_u)$ denotes the $K$ token-history slots, $t(\tilde{H}_u)$ is the titles line, and $P_{\mathrm{judge}}$ is a fixed judgment suffix. The token-history slots are computed once from $C_u$ and shared by all $N$ prompts, and the ranking is $\operatorname{argsort}_c\, s_c$. A logit readout avoids the parsing failures of generative rankers. The complete rendered prompts, one masked co-training event and one test instance, are listed verbatim in the technical appendix.

\textbf{Instruction placement.}
The instruction competes with the collaborative tokens for attention, so its position $\pi$ is consequential. We allow three slots (early, middle, or late in the prompt; slot definitions in the technical appendix) and choose one per dataset with a zero-shot rule card: an LLM, shown a compact card of training-set statistics only, predicts the slot. The position is \emph{predicted, not searched}; no test feedback informs it.

\section{Experiments}
\label{sec:experiments}

\subsection{Experimental Setup}
\label{sec:setup}

\textbf{Datasets and Evaluation Protocol.}
We use the four instruction-grounded recommendation domains of InstructRec \citep{xu2025iagent}: Amazon Book (Books), Goodreads, Amazon MovieTV (MovieTV), and Yelp, with the released user instructions and splits; Table~\ref{tab:datasets} summarizes their statistics. Interactions are ordered by timestamp per user; the last item is the test target and the second-to-last is held out for validation (leave-one-out). Each test instance ranks $N{=}10$ candidates: the held-out ground truth plus nine negatives sampled uniformly outside the user's entire sequence (seed 42). This candidate-list size follows MemRec's evaluation setting \citep{memrec2026}. We report Hit@$K$ (H@$K$) for $K \in \{1,3,5\}$ and NDCG@$K$ (N@$K$) for $K \in \{3,5\}$ on the full test sets.

\begin{table}[t]
\centering
\small
\begin{tabular*}{\columnwidth}{@{}l@{\extracolsep{\fill}}rrrrr@{}}
\toprule
Dataset & \#Users & \#Items & \#Inter. & $\bar{L}_u$ & Density \\
\midrule
Books & 7.4K & 120.9K & 207.8K & 28.2 & $2.33{\times}10^{-4}$ \\
Goodreads & 11.7K & 57.4K & 618.3K & 52.7 & $9.19{\times}10^{-4}$ \\
MovieTV & 5.6K & 29.0K & 79.7K & 14.1 & $4.87{\times}10^{-4}$ \\
Yelp & 3.0K & 31.6K & 63.1K & 21.4 & $6.77{\times}10^{-4}$ \\
\bottomrule
\end{tabular*}
\caption{Dataset statistics; $\bar{L}_u$ is the average history length per user.}
\label{tab:datasets}
\end{table}

\textbf{Baselines.}
Non-agentic recommenders comprise LightGCN \citep{he2020lightgcn}, SASRec \citep{kang2018sasrec}, and the text-to-text P5 \citep{geng2022p5}. The LLM-based baselines are Vanilla LLM \citep{liu2023chatgpt,memrec2026}, with no explicit memory, and the text-memory agents of \S\ref{sec:related}: iAgent and i$^2$Agent \citep{xu2025iagent}, AgentCF \citep{zhang2024agentcf}, and MemRec \citep{memrec2026}, the strongest under this protocol. All baselines are our reproductions, and the agentic ones run on the same LLM stack as our scorer.

\textbf{Implementation.}
The scorer is Qwen2.5-7B-Instruct \citep{qwen25techreport} (bfloat16, frozen) with a LoRA adapter of rank $r{=}4$ on the attention projections. The gated projector maps the $d{=}64$ states into the LLM embedding space, and the states come from LightGCN trained on the train-only interaction graph. The memory schema carries $K{=}5$ target-aware token-history slots plus a titles line for the same $K$ events as a light textual complement. Instruction positions follow the zero-shot rule card: early for Goodreads, middle for MovieTV, late for Yelp and Books. The user's textual profile is distilled once offline by the same LLM backbone; architecture details are listed in the technical appendix.

\textbf{Training.}
Co-training runs 2 epochs over enumerated events (from each user's fourth interaction onward); Goodreads/MovieTV/Books cap events at 25{,}000 per epoch with per-epoch resampling, and Yelp uses its full stream ($\approx$48k). The projector trains at a learning rate of $10^{-4}$, the LoRA adapter at one twentieth of it, with batch size 16. Checkpoints are selected on a fixed 400-user subsample of the validation split (drawn once, seed 4242); the remaining hyperparameters and the selection protocol are detailed in the technical appendix. All experiments run on a single NVIDIA A800 (80\,GB) GPU with PyTorch 2.11, Transformers 5.13, and PEFT 0.19.

\subsection{Overall Performance (RQ1)}
\label{sec:main-results}

\begin{table*}[t]
\centering
\small
\begin{tabular*}{\textwidth}{@{}l@{\extracolsep{\fill}} ccccc ccccc@{}}
\toprule
& \multicolumn{5}{c}{Books} & \multicolumn{5}{c}{Goodreads} \\
\cmidrule(lr){2-6} \cmidrule(lr){7-11}
Method & H@1$\uparrow$ & H@3$\uparrow$ & H@5$\uparrow$ & N@3$\uparrow$ & N@5$\uparrow$ & H@1$\uparrow$ & H@3$\uparrow$ & H@5$\uparrow$ & N@3$\uparrow$ & N@5$\uparrow$ \\
\midrule
LightGCN      & 0.1693 & 0.3263 & 0.5551 & 0.2572 & 0.3502 & \underline{0.8132} & \textbf{0.9344} & \textbf{0.9588} & \textbf{0.8857} & \textbf{0.8959} \\
SASRec        & 0.2169 & 0.4915 & 0.6824 & 0.3732 & 0.4515 & \textbf{0.8149} & \underline{0.9246} & 0.9510 & \underline{0.8804} & \underline{0.8914} \\
P5            & 0.1391 & 0.3336 & 0.5219 & 0.2489 & 0.3259 & 0.5848 & 0.7533 & 0.8373 & 0.6832 & 0.7177 \\
\midrule
Vanilla LLM   & 0.4106 & 0.6508 & \underline{0.7888} & 0.5488 & 0.6053 & 0.2082 & 0.4590 & 0.6480 & 0.3511 & 0.4286 \\
iAgent & 0.3645 & 0.5372 & 0.6833 & 0.4641 & 0.5240 & 0.2201 & 0.4905 & 0.6771 & 0.3750 & 0.4515 \\
AgentCF       & 0.4219 & 0.6360 & 0.7668 & 0.5453 & 0.5990 & 0.3972 & 0.6445 & 0.7460 & 0.5412 & 0.5829 \\
i$^2$Agent & 0.4200 & 0.6017 & 0.7341 & 0.5251 & 0.5792 & 0.2996 & 0.5515 & 0.7256 & 0.4439 & 0.5152 \\
MemRec        & \textbf{0.4837} & \underline{0.6706} & 0.7789 & \underline{0.5913} & \underline{0.6358} & 0.3087 & 0.5681 & 0.7274 & 0.4583 & 0.5238 \\
\midrule
CoVeMem       & \underline{0.4732} & \textbf{0.7157} & \textbf{0.8447} & \textbf{0.6135} & \textbf{0.6665} & 0.7730 & 0.9223 & \underline{0.9575} & 0.8622 & 0.8767 \\
\bottomrule
\end{tabular*}

\vspace{6pt}

\begin{tabular*}{\textwidth}{@{}l@{\extracolsep{\fill}} ccccc ccccc@{}}
\toprule
& \multicolumn{5}{c}{MovieTV} & \multicolumn{5}{c}{Yelp} \\
\cmidrule(lr){2-6} \cmidrule(lr){7-11}
Method & H@1$\uparrow$ & H@3$\uparrow$ & H@5$\uparrow$ & N@3$\uparrow$ & N@5$\uparrow$ & H@1$\uparrow$ & H@3$\uparrow$ & H@5$\uparrow$ & N@3$\uparrow$ & N@5$\uparrow$ \\
\midrule
LightGCN      & 0.3679 & 0.5666 & 0.6867 & 0.4835 & 0.5326 & 0.3844 & 0.6244 & 0.8037 & 0.5236 & 0.5971 \\
SASRec        & 0.3914 & 0.5553 & 0.6759 & 0.4865 & 0.5359 & 0.2702 & 0.4780 & 0.6115 & 0.3901 & 0.4451 \\
P5            & 0.4815 & 0.6752 & 0.8005 & 0.5931 & 0.6447 & 0.1563 & 0.3749 & 0.5631 & 0.2809 & 0.3580 \\
\midrule
Vanilla LLM   & 0.3850 & 0.6148 & 0.7435 & 0.5183 & 0.5710 & 0.4224 & \underline{0.6373} & 0.7644 & 0.5476 & 0.5999 \\
iAgent & 0.3437 & 0.5771 & 0.7282 & 0.4784 & 0.5404 & 0.3313 & 0.5295 & 0.6804 & 0.4467 & 0.5085 \\
AgentCF       & 0.4020 & 0.6429 & 0.7745 & 0.5411 & 0.5952 & 0.3112 & 0.5471 & 0.7064 & 0.4461 & 0.5117 \\
i$^2$Agent & 0.4549 & 0.6718 & 0.8070 & 0.5803 & 0.6358 & 0.3418 & 0.5392 & 0.6863 & 0.4555 & 0.5157 \\
MemRec        & \underline{0.5371} & \underline{0.7578} & \underline{0.8494} & \underline{0.6663} & \underline{0.7039} & \underline{0.5173} & \textbf{0.7424} & \underline{0.8275} & \underline{0.6488} & \underline{0.6839} \\
\midrule
CoVeMem       & \textbf{0.5799} & \textbf{0.8042} & \textbf{0.8929} & \textbf{0.7108} & \textbf{0.7473} & \textbf{0.5400} & \textbf{0.7424} & \textbf{0.8298} & \textbf{0.6574} & \textbf{0.6934} \\
\bottomrule
\end{tabular*}
\caption{Main results. Best and second-best results are in bold and underlined, respectively. All improvements are statistically significant ($p < 0.05$).}
\label{tab:main}

\end{table*}

\begin{figure*}[t]
  \centering
  \includegraphics[width=0.97\textwidth]{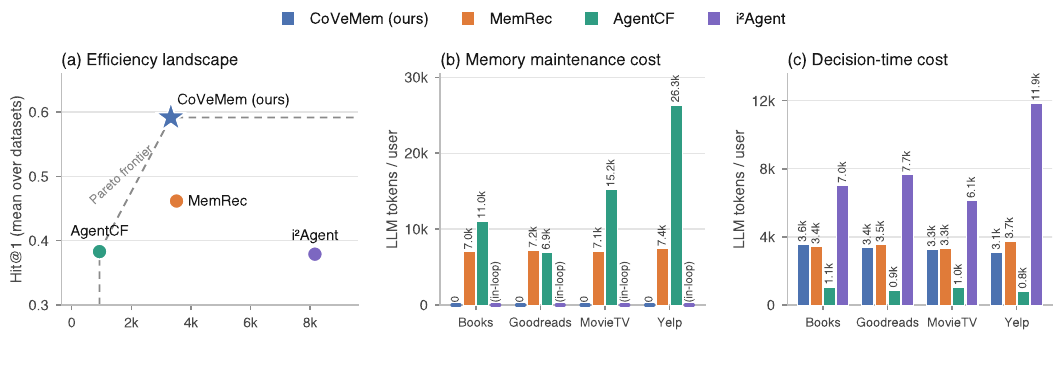}
  \caption{Memory efficiency (RQ4). (a) Mean Hit@1 versus decision-time LLM tokens per user (dashed line: Pareto frontier). (b) Memory-maintenance and (c) decision-time cost per user; i$^2$Agent maintains within its decision calls, so its maintenance appears in (c).}
  \label{fig:efficiency}
\end{figure*}

Table~\ref{tab:main} compares CoVeMem with all baselines on the four datasets. Against MemRec, the strongest text-memory agent, CoVeMem matches or exceeds on 19 of the 20 metric cells. The advantage is largest on Goodreads, while MovieTV shows the same direction across all five metrics at smaller margins. All remaining text-memory agents fall below CoVeMem on every metric.

Goodreads is popularity-dominated: ordering candidates by training-set popularity alone reaches a Hit@1 of 0.6727, showing that population-level interaction statistics provide a strong ranking signal. CoVeMem's collaborative vector memory encodes catalog-wide interaction structure in its graph-trained memory bank and exposes it to the LLM through retrieved states, whereas text memory retains only selected facets in serialized form. With this collaborative evidence, CoVeMem scores above the popularity-only ranker and recovers over ninety percent of the Hit@1 gap between Vanilla LLM and the strongest conventional recommender.

Books and Yelp have lower average interactions per item, so each item state is estimated from less direct collaborative evidence. In this regime, textual semantics play a larger role in the decision. CoVeMem preserves the LLM's ability to exploit this textual evidence while adding collaborative information as a complement: it leads on Yelp and stays on par on Books.

\subsection{Memory Ablation Study (RQ2)}
\label{sec:soft-token-ablation}

\begin{table}[t]
\centering
\small
\setlength{\tabcolsep}{2.0pt}\begin{tabular*}{\columnwidth}{@{}ll@{\extracolsep{\fill}}ccccc@{}}
\toprule
Dataset & Memory & H@1 & H@3 & H@5 & N@3 & N@5 \\
\midrule
\multirow{4}{*}{Goodreads} & \textsc{Text-Recent} & 0.3446 & 0.6260 & 0.7822 & 0.5056 & 0.5697 \\
 & \textsc{Text-TA} & 0.3567 & 0.8649 & 0.9316 & 0.6643 & 0.6921 \\
 & w/o LoRA & 0.1001 & 0.3229 & 0.5589 & 0.2261 & 0.3226 \\
 & CoVeMem & \textbf{0.7730} & \textbf{0.9223} & \textbf{0.9575} & \textbf{0.8622} & \textbf{0.8767} \\
\midrule
\multirow{4}{*}{Yelp} & \textsc{Text-Recent} & 0.3614 & 0.5953 & 0.7336 & 0.4960 & 0.5531 \\
 & \textsc{Text-TA} & 0.4339 & 0.7041 & 0.8119 & 0.5921 & 0.6367 \\
 & w/o LoRA & 0.4180 & 0.6427 & 0.7729 & 0.5479 & 0.6014 \\
 & CoVeMem & \textbf{0.5400} & \textbf{0.7424} & \textbf{0.8298} & \textbf{0.6574} & \textbf{0.6934} \\
\bottomrule
\end{tabular*}
\caption{Memory component ablation. Best results are in bold.}
\label{tab:carrier}
\end{table}

We ablate the collaborative vector memory and the LoRA component of the parametric memory reader. \textsc{Text-TA} and \textsc{Text-Recent} disable the collaborative vector memory, differing in history-title selection: target-aware (Section~\ref{sec:read}) versus recency. w/o LoRA freezes the adapter at zero initialization while the gated projector and state injection follow the full recipe. Table~\ref{tab:carrier} examines Goodreads and Yelp as diagnostic cases, where RQ1 shows the clearest separation between vector and text memory.

The text-only variants fall to the text-memory-agent range of Table~\ref{tab:main}. \textsc{Text-TA} generally outperforms \textsc{Text-Recent}, so target-aware selection helps even without injected states. Most of the full model's Hit@1 advantage nevertheless persists under either selection rule, so it comes primarily from the injected states. w/o LoRA drops to the Vanilla LLM level on Yelp and below it on Goodreads, reaching chance level (Table~\ref{tab:main}). Because attention is never adapted to the injected states, the model cannot read them and they actively interfere with the textual evidence, confirming that the LoRA adapter is necessary for the collaborative vector memory to be effective.

\subsection{Collaborative Backbone Study (RQ3)}
\label{sec:backbone}

\begin{table}[t]
\centering
\small
\setlength{\tabcolsep}{4pt}\begin{tabular*}{\columnwidth}{@{}l@{\extracolsep{\fill}}ccccc@{}}
\toprule
Backbone & H@1 & H@3 & H@5 & N@3 & N@5 \\
\midrule
Text-only memory & 0.4339 & 0.7041 & 0.8119 & 0.5921 & 0.6367 \\
\midrule
SASRec & 0.4369 & 0.6668 & 0.7956 & 0.5692 & 0.6222 \\
BPR-MF & 0.4736 & 0.6881 & 0.7983 & 0.5978 & 0.6430 \\
GRU4Rec & 0.4990 & 0.7153 & \underline{0.8325} & 0.6244 & 0.6728 \\
LightGCN (default) & \underline{0.5400} & \underline{0.7424} & 0.8298 & \underline{0.6574} & \underline{0.6934} \\
SVD++ & \textbf{0.5569} & \textbf{0.7763} & \textbf{0.8644} & \textbf{0.6856} & \textbf{0.7217} \\
\bottomrule
\end{tabular*}
\caption{Collaborative-backbone study on Yelp, with the pipeline retrained for each backbone. Best and second-best backbone results are in bold and underlined, respectively.}
\label{tab:backbone}
\end{table}

The previous section attributed the gains to the injected states; we now ask whether they depend on the specific model that produced those states. The memory bank is modular: any conventional recommender that yields user and item states over the training interactions can serve as its collaborative backbone. We therefore replace the Yelp memory bank with states from five collaborative backbones spanning three families: matrix factorization (BPR-MF \citep{rendle2009bpr}, SVD++ \citep{koren2008svdpp}), graph propagation (LightGCN \citep{he2020lightgcn}), and sequential encoding (GRU4Rec \citep{hidasi2016gru4rec}, SASRec \citep{kang2018sasrec}). We then rerun the identical pipeline end to end: same alignment, co-training, candidates, and readout; imported state tables are rescaled where their own scale degenerates, as detailed in the technical appendix.

Table~\ref{tab:backbone} shows that the pipeline transfers across collaborative backbones: all five train successfully, and the three strongest outperform the text-only variant of RQ2 (included for reference) on every metric. At the same time, the wide spread indicates that the quality of the memory bank directly determines how much the agent benefits from the memory. The strongest collaborative backbone surpasses the default LightGCN on every metric, while the weakest shows no clear advantage over the text-only variant: the vector memory is only as effective as the collaborative backbone behind it.

\subsection{Memory Maintenance Efficiency (RQ4)}
\label{sec:efficiency}

The preceding sections established what the vector memory contributes; we now examine what it costs. We measure two system-specific costs beyond the shared static profile: a memory-maintenance cost, the tokens of the generative LLM calls that construct or update each method's additional memory, and a decision-time cost, the tokens spent scoring candidates. We measure both on all four datasets from recorded token counts rather than estimates. Figure~\ref{fig:efficiency} reports both costs.

The three text-memory agents distribute these costs differently. MemRec builds its memory with the main LLM, writing facets by reflection and propagating them to neighboring users along the interaction graph; constructing the memory alone consumes tens of millions of LLM tokens per dataset before any candidate is scored. AgentCF co-adjusts the user's and the item's text memories after every training interaction, which gives it the highest average maintenance cost in Figure~\ref{fig:efficiency}b.
i$^2$Agent instead performs maintenance at inference time: it builds nothing offline, but updates its profile, knowledge, and interest components through separate LLM stages at every decision. In all three cases, maintaining the memory requires recurrent generative LLM calls, which add token cost and serial latency to the memory lifecycle.

Under this accounting boundary, CoVeMem incurs zero LLM tokens for memory maintenance. Its persistent components are constructed once offline (\S\ref{sec:setup}), and no generative LLM call subsequently updates the memory. With a non-generative yes/no readout, its per-user token cost therefore consists only of decision-time inputs; this total is the lowest of the four systems on every dataset and comparable to MemRec's decision-time cost alone. In Figure~\ref{fig:efficiency}(a), CoVeMem occupies the high-accuracy end of the Pareto frontier.

\section{Conclusion}
\label{sec:conclusion}

We presented CoVeMem, whose collaborative vector memory forms the collaborative component of an agentic recommender's complete memory: target-aware retrieval selects frozen collaborative states into the LLM context, and contrastive alignment with masked co-training teaches the model to read them. Across four benchmarks, CoVeMem matches or exceeds the strongest text-memory agent while requiring zero additional memory-maintenance calls beyond the shared static profile. The gains are largest where collaborative statistics decide and text-memory agents are weakest: evidence arrives in its native form rather than as sentences, preserving the LLM's semantic strengths while adding information from conventional recommenders. More broadly, the memory carrier emerges as a design axis of its own. Prior work enriched what a text memory writes; the carrier decides what evidence can reach a decision at all. Once the memory takes gradients, it stops being an artifact the agent must write and becomes a component it can train. The full interaction history, beyond what text can absorb, becomes available as event-level supervision, and any conventional recommender can supply the memory's collaborative information. The current memory uses one collaborative backbone; composing several backbones into a higher-quality memory remains an interesting direction.

\bibliography{references}

\lstset{%
	basicstyle={\rmfamily},%
	numbers=none,xleftmargin=0pt,%
	aboveskip=0pt,belowskip=0pt,%
	showstringspaces=false,tabsize=2,breaklines=true,columns=fullflexible,keepspaces=true}
\appendix

\noindent This appendix specifies CoVeMem's instruction-position rule, textual profile construction, training configuration, checkpoint selection, and exact prompt artifacts.

\section{Implementation Details}

\subsection{Instruction Positions}
\label{app:positions}

The zero-shot rule card of the main paper chooses among three instruction slots: \emph{early}, before the user context; \emph{middle}, between the memory and the candidate; and \emph{late}, next to the judgment suffix. Listing~\ref{lst:prompt} shows the early position as deployed on Goodreads.

\textbf{Instruction-Position Rule Card.}
Let $N_{\mathrm{int}}$ and $N_{\mathrm{item}}$ denote the numbers of training interactions and items, respectively, and let $N_{\mathrm{head}}$ denote the number of training interactions involving the most popular 1\% of items. The generator receives two train-only statistics: the interaction density $d$ and the concentration ratio $q$,
\[
d=\frac{N_{\mathrm{int}}}{N_{\mathrm{item}}},
\qquad
q=\frac{N_{\mathrm{head}}}{N_{\mathrm{int}}}.
\]
Given the mechanism description and these train-only statistics, the LLM returned the following rule without access to validation or test results: assign \emph{early} when $d\geq5$ or $q\geq25\%$, \emph{middle} when neither early threshold is met but $d\geq2.5$ or $q\geq12\%$, and \emph{late} otherwise. In the same output, it assigns Goodreads ($d{=}10.54$, $q{=}39.2\%$) to the early position, MovieTV ($d{=}2.61$, $q{=}13.0\%$) to the middle position, and both Yelp ($d{=}1.96$, $q{=}9.8\%$) and Books ($d{=}1.74$, $q{=}9.8\%$) to the late position.

\subsection{Textual Profile Construction}
\label{app:profiles}

Each user's textual profile is constructed exclusively from train-split interactions, reviews, and item metadata. The input contains at most the 12 most recent training items; each title and category field is truncated to 60 characters, and each review to 300 characters. Qwen2.5-7B-Instruct generates the profile with temperature 0 and a maximum of 160 tokens. Listing~\ref{lst:profileprompt} reproduces the Yelp generation template.

The generated profile is normalized and stored once offline; it is not updated during training or evaluation. When rendered in the prompt, it is converted to one line and truncated uniformly to 220 characters. The code supplement provides the generation script and domain-matched templates for reproducing all profiles.

\subsection{Training Configuration}
\label{app:hparams}

The LoRA adapter uses rank $r{=}4$, $\alpha{=}8$, and dropout $0.05$ on all four attention projections ($W_q$, $W_k$, $W_v$, $W_o$). The gated projector maps $d{=}64$ states through two gated blocks with hidden widths 64, 256, 1024, and 3584, followed by a LayerNorm output; token-history slots are zero-padded when the history holds fewer than $K$ items. The LightGCN bank uses 3 propagation layers and BPR training for 50 epochs on the train-only graph (seed 42). Alignment uses temperature $\tau{=}0.07$. Co-training masks candidate titles with probability $p{=}0.5$ and applies a hinge margin $m{=}1.0$. The gate warms up over $T_g{=}60$ steps with soft-token dropout $0.1$. We set the projector learning rate to $10^{-4}$, the LoRA learning rate to $5{\times}10^{-6}$, weight decay to $0.01$, and gradient clipping to $1.0$. The batch size is 16 without gradient accumulation. In the collaborative-backbone study, imported state tables whose embedding norms deviate substantially from the default bank are rescaled (factor $3.5$ for user embeddings, $0.8$ for item embeddings).

All reported experiments ran under Ubuntu 22.04.5 LTS on an Intel Xeon Gold 6348 host with a single NVIDIA A800 (80\,GB) GPU, using the software stack specified in the main paper.

\subsection{Checkpoint Selection}
\label{app:ckpt}

After each co-training epoch, we retain the checkpoint with the best ranking on a fixed 400-user validation subsample (drawn once; seed 4242), using the same pointwise yes/no readout as at test time.

\section{Prompt Artifacts}
\label{app:prompt}

\subsection{Instruction-Position Rule Generator}

Listings~\ref{lst:ruleprompt-context} and~\ref{lst:ruleprompt-task} reproduce the complete input to the zero-shot rule generator.

\begin{listing*}[t]
\caption{Rule-generator input, part 1 of 2: role, prompt context, and position mechanism.}
\label{lst:ruleprompt-context}
\begin{lstlisting}
ROLE: You are the offline rule generator of a recommendation system (the exact analog of MemRec's LLM-as-Rule-Generator, its Appendix A.4). You are given ONLY train-split domain statistics and a mechanism description. You must output an interpretable, per-domain configuration rule. You have NO access to any evaluation results.

SYSTEM CONTEXT: A frozen-backbone LLM scores each candidate with a pointwise prompt that contains, in order: [user line, (optional position A) the user live instruction, textual profile, historical titles, a collaborative user memory token, the candidate line with a collaborative item memory token, (optional position C) the instruction, scoring task]. Position B places the instruction between the user memory token and the candidate line. The design choice per domain: place the live instruction EARLY (position A), MIDDLE (B), or LATE (C).

MECHANISM (established, not dataset-specific): tokens closer to the final scoring position receive higher effective weight (recency). Placing the instruction LATE amplifies its influence on the verdict but displaces the candidate line and collaborative memory tokens one block away from the decision point, diluting collaborative evidence. EARLY does the opposite: collaborative tokens sit closest to the decision, the instruction acts as distant framing. MIDDLE is intermediate: instruction gains recency over EARLY while the candidate/token block keeps decision-point adjacency.
\end{lstlisting}
\end{listing*}

\begin{listing*}[t]
\caption{Rule-generator input, part 2 of 2: train-only statistics and required output.}
\label{lst:ruleprompt-task}
\begin{lstlisting}

TRAIN-ONLY DOMAIN STATISTICS:
- yelp (local businesses): 2950 users, 57242 interactions, 29263 items; 19.4 inter/user; 1.96 users/item; top-1% item share 9.8%.
- goodreads (books, social reading): 11734 users, 594862 interactions, 56432 items; 50.7 inter/user; 10.54 users/item; top-1% item share 39.2%.
- movietv (films/TV): 5649 users, 68439 interactions, 26196 items; 12.1 inter/user; 2.61 users/item; top-1% share 13.0%.
- books (amazon books): 7377 users, 193005 interactions, 111084 items; 26.2 inter/user; 1.74 users/item; top-1% share 9.8%.

TASK: Reason about how collaborative-signal density (users/item, popularity concentration) determines the reliability of the collaborative memory tokens relative to the live instruction in each domain, then output:
1. A general rule (2-4 sentences, interpretable, thresholds allowed) mapping train statistics -> instruction position.
2. The per-domain assignment (yelp/goodreads/movietv/books -> early|middle|late) with one-line rationales.
Output as a compact rule card. Do not hedge with multiple options per domain; commit to one position each.
\end{lstlisting}
\end{listing*}

\subsection{Textual Profile Generator}

Listing~\ref{lst:profileprompt} shows the Yelp user-message template. The MovieTV and Books/Goodreads variants preserve its instruction and output constraints while substituting domain-specific evidence labels and preference facets; their exact templates are included in the code supplement. Here, \texttt{\{training\_items\}} is replaced by up to 12 lines of train-only evidence. A line is rendered as \texttt{- \{title\} (\{categories\})} and, when a review is available, followed by \texttt{| the user wrote: ``\{review\}''}.

\begin{listing}[t]
\caption{User-message template for offline textual-profile construction (Yelp).}
\label{lst:profileprompt}
\begin{lstlisting}
[Yelp]
Summarize this user's dining/business preferences in 2-3 concise sentences, based on the places they visited and what they wrote. Mention cuisines, atmosphere, price sensitivity, and recurring themes. Be specific and objective; do not invent preferences not supported by the evidence; do not list raw item names.

User's visited places and their own reviews (train history only):
{training_items}

User preference profile:
\end{lstlisting}
\end{listing}

\subsection{Training and Evaluation Prompts}

Listings~\ref{lst:trainprompt} and~\ref{lst:prompt} show the pipeline's two prompt forms for the same Goodreads user. Listing~\ref{lst:trainprompt} is a listwise co-training event with shuffled candidate order and no instruction. Each candidate is independently masked with probability $p$; masked candidates retain only their item tokens. In this draw, the positive, option~[J], is masked, so the loss is restricted to the masked subset. Listing~\ref{lst:prompt} is the per-candidate yes/no test query, complementing Figure~2 in the main paper. Both prompts instantiate the same memory schema but differ in two respects. First, the user-token line appears in a different position, but lookup by token ID ensures that each vector is injected at its corresponding placeholder despite the different order. Second, the selected history differs because target-aware selection depends on each decision's candidate set.

\begin{listing*}[t]
\caption{One listwise co-training event as rendered (Goodreads; candidate masking $p{=}0.5$, one draw shown; no instruction in training). Bold markers are the injected soft-token positions.}
\label{lst:trainprompt}
\begin{lstlisting}[escapeinside={(*@}{@*)}]
**Target User:** User 0
**User Preferences (Static Ego Memory):**  1. Based on the user's reviews, it appears they have a strong interest in young adult fiction, particularly series such as The Mortal Instruments, Vampire Academy, and Divergent. They seem to enjoy books with complex cha...
**Collaborative User Soft Token:** (*@\textbf{<USER\_EMB>}@*)
**Relevant Memories:** (*@\textbf{<ITEM\_EMB>}@*) (*@\textbf{<ITEM\_EMB>}@*) (*@\textbf{<ITEM\_EMB>}@*) (*@\textbf{<ITEM\_EMB>}@*) (*@\textbf{<ITEM\_EMB>}@*)
**Recent Interactions (oldest to newest):**
  - Divergent (Divergent, #1)
  - Hush, Hush (Hush, Hush, #1)
  - Clockwork Angel (The Infernal Devices, #1)
  - Insurgent (Divergent, #2)
  - The Fault in Our Stars
**Candidate Options:**
[A] (*@\textbf{<ITEM\_EMB>}@*) The Discovery (Pyxis, #1)
[B] (*@\textbf{<ITEM\_EMB>}@*) [MASKED]
[C] (*@\textbf{<ITEM\_EMB>}@*) [MASKED]
[D] (*@\textbf{<ITEM\_EMB>}@*) [MASKED]
[E] (*@\textbf{<ITEM\_EMB>}@*) Thick as Thieves (The Queen's Thief, #5)
[F] (*@\textbf{<ITEM\_EMB>}@*) America Through the Lens: Photographers Who Changed the Nation
[G] (*@\textbf{<ITEM\_EMB>}@*) Mystery at Chances Hill
[H] (*@\textbf{<ITEM\_EMB>}@*) Entangled (Entangled, #1)
[I] (*@\textbf{<ITEM\_EMB>}@*) [MASKED]
[J] (*@\textbf{<ITEM\_EMB>}@*) [MASKED]

**Scoring Task:**
Select the single option that is the best held-out match for the user.
\end{lstlisting}
\end{listing*}

\begin{listing*}[t]
\caption{A rendered per-candidate scoring prompt (Goodreads; early instruction position; the shown candidate is the held-out positive). Bold markers are the injected soft-token positions: the five token-history slots, the user token, and the candidate's item token.}
\label{lst:prompt}
\begin{lstlisting}[escapeinside={(*@}{@*)}]
**Target User:** User 0
**User's Current Request:**
I'm inclined to explore the works of Johnson, as my appreciation for New Adult romance is piqued by their creative expressions and compelling narratives.
**User Preferences (Static Ego Memory):**  1. Based on the user's reviews, it appears they have a strong interest in young adult fiction, particularly series such as The Mortal Instruments, Vampire Academy, and Divergent. They seem to enjoy books with complex cha...
**Relevant Memories:** (*@\textbf{<ITEM\_EMB>}@*) (*@\textbf{<ITEM\_EMB>}@*) (*@\textbf{<ITEM\_EMB>}@*) (*@\textbf{<ITEM\_EMB>}@*) (*@\textbf{<ITEM\_EMB>}@*)
**Recent Relevant Interactions:** Hush, Hush (Hush, Hush, #1); Beautiful Creatures (Caster Chronicles, #1); City of Lost Souls (The Mortal Instruments, #5); Insurgent (Divergent, #2); The Fault in Our Stars
**Collaborative User Soft Token:** (*@\textbf{<USER\_EMB>}@*)
**Candidate Item Memory (8 of 10):**  - (*@\textbf{<ITEM\_EMB>}@*) Item 26574 (Where We Fell): Oliver Bishop is having a seriously bad day. With one diagnosis, his life sudden
**Scoring Task:**
Score only the candidate above. Answer Yes if this item is the best held-out match for the user among the current 10 candidates; otherwise answer No.
Answer:
\end{lstlisting}
\end{listing*}

\end{document}